\documentclass[journal]{IEEEtran}

\usepackage{cite}

\usepackage{tikz}

\usepackage[english]{babel}

\usepackage{etoolbox}
\apptocmd{\thebibliography}{\raggedright}{}{}

\ifCLASSINFOpdf
\else
\fi
\usepackage[cmex10]{amsmath}

\usepackage{array}

\usepackage{multirow}

\usepackage{graphicx}

\usepackage{color}
\usepackage{xcolor}
\usepackage{soul}
\usepackage{caption}

\usepackage{stfloats}
\usepackage{amsmath,amsfonts,amssymb,amscd}
\usepackage{mathrsfs}
\usepackage{graphicx,epsfig}

\usepackage{cases}
\usepackage{multirow}

\usepackage{lscape}
\usepackage{verbatim}
\usepackage{epstopdf}
\usepackage{balance}

\usepackage{psfrag}
\psfrag{J}{$\mathbb{E}$}

\usepackage{bbm}

\usepackage{lipsum}

\usepackage{bbold}

\begin{document}
%
\title{Gaussian Process Regression for Probabilistic Short-term Solar Output Forecast}

\author{\IEEEauthorblockA{\large{Fatemeh Najibi, Dimitra Apostolopoulou, and Eduardo Alonso}}

\thanks{F. Najibi and E. Alonso are with the Artificial Intelligence Research Centre in the Department of Computer Science and D. Apostolopoulou with the Department of Electrical and Electronic Engineering  at City, University of London, London, UK EC1V 0HB. E-mail: \texttt{\{Fatemeh.Najibi, Dimitra.Apostolopoulou, E.Alonso\}@city.ac.uk}}}

\maketitle

\begin{abstract}
With increasing concerns of climate change, renewable resources such as photovoltaic (PV) have gained popularity as a means of energy generation. The smooth integration of such resources in power system operations is enabled by accurate forecasting mechanisms that address their inherent intermittency and variability. This paper proposes a probabilistic framework to predict short-term PV output taking into account the uncertainty of weather. To this end, we make use of datasets that comprise of power output and meteorological data such as irradiance, temperature, zenith, and azimuth. First, we categorise the data into four groups based on solar output and time by using k-means clustering. Next, a correlation study is performed to choose the weather features which affect solar output to a greater extent. Finally, we determine a function that relates the aforementioned selected features with solar output by using Gaussian Process Regression and Mat\'ern 5/2 as a kernel function. We validate our method with five solar generation plants in different locations and compare the results with existing methodologies. More specifically, in order to test  the proposed model, two different methods are used: (i) a 5-fold cross validation; and (ii) holding out 30 random days as test data. To confirm the model accuracy, we apply our framework 30 independent times on each of the four clusters. The average error follows a normal distribution, and with 95\%  confidence level, it takes values between $-$1.6\% to 1.4\%. 
\end{abstract}
\hfill \break
\begin{IEEEkeywords}
Short-term forecasting, photovoltaic, Gaussian Processes Regression, k-means, feature selection.
\end{IEEEkeywords}



%

\section{Introduction}

Over the past years many countries have opted to integrate solar energy in the grid in order to increase the penetration of environmental friendly resources~\cite{doi:10.1177/0144598716650552}. For instance, Japan, China, Germany, USA, and UK are able to meet 80\% of their demand from solar generation; and the total installed solar energy capacity at the end of 2018 was more than 500 GW~\cite{pinsonbook}. However, the inherent uncertainty of photovoltaic (PV) energy, due to, e.g., irradiance, temperature and cloud conditions, makes its smooth integration in power system operations a formidable challenge. More specifically, the intermittency of solar generation might cause issues in system stability, power balance and frequency response, and reactive power generation (see, e.g.,~\cite{SHAH20151423,6807836}). In this regard, building accurate forecast models of solar generation is of vital importance.  
 
PV output forecasting may be classified into four groups based on the approach used to model solar panels and weather behaviour, namely: (i) statistical methods; (ii) Artificial Intelligence (AI); (iii) physical models; and (iv) hybrid approaches~\cite{7570246}. Statistical approaches are based on available historical measured meteorological and PV output data as well as numerical weather forecasts. AI methods use machine learning techniques such as Artificial Neural Networks (ANNs) to capture the non-linear relationship between weather data and solar output and construct a probabilistic model~\cite{7280574}. These methods may be classified in group (i) above if their performance is judged by statistical metrics~\cite{BEHERA2018428}. Physical models focus on numerical weather forecasts and the use of satellite images to predict weather parameters such as solar irradiation as input to a PV model to determine the solar generation output. Last, hybrid models combine the aforementioned approaches.

There are several advantages and disadvantages associated with each group of methods. For instance, in physical models where a detailed description of the panels based on the single diode model is used, the stochastic nature of weather data is neglected (see, e.g., \cite{Najibi}). Moreover, the output is based on a PV datasheet, therefore the partial failure and down time of a PV plant are not considered. As such, physical models usually have less accuracy in their forecasts compared against AI algorithms. Other studies pivot around statistical approaches or hybrid models that incorporate machine learning and statistical techniques. In~\cite{7565620} a probabilistic forecast model is proposed as a linear programming model. The authors used Extreme Machine Learning (ELM) and quantile regression to efficiently develop a statistical approach to generate a confidence interval on the forecasted power generation. In~\cite{7592462, Golestaneh}, different distribution functions are combined to predict a 15-minute ahead probability distribution function of PV output based on a higher-order Markov chain. This method has been recently proved to improve generalisation in comparison to standard back-propagation \cite{4012031}. Although a plethora of contemporary studies have focused on ANNs and Support Vector Regression approaches in the context of forecasting~\cite{VOYANT2017569}, other machine learning techniques such as regression trees can also be used based on available historical data.  According to~\cite{PEDRO20122017}, which discusses the assessment of different forecasting techniques, most ANNs and persistence models disregard the uncertainty provoked by the random behaviour of meteorological data. On the other hand, regression techniques incorporate uncertainty and are able to build a probabilistic forecast model. For example in~\cite{6168891, 6348273}, the authors utilised a Support Vector Machine (SVM) to predict PV output based on different meteorological conditions. 

Among all the approaches used to predict solar output, Gaussian Process Regression (GPR) is one of the most powerful due to its flexibility to be applied on a wide range of time-series data~\cite{gprrr}. GPR is a unique method for modelling uncertainty in a probabilistic framework setup~\cite{LI2014325}. In modelling weather forecast, the uncertainty of input attributes are taken into account by using GPR which treats input data as random points with an unknown distribution function. Therefore, the uncertainties are reflected into the output forecast with a specific confidence interval. GPR is based on Bayesian statistics, which help us model and quantify uncertainty in the parameters. Moreover, the non-linear relationship between solar output and meteorological weather parameters can be explicitly modelled by using an appropriate kernel function~\cite{HEO20127}. In comparison to other methods, GPR is more efficient for prediction in time-series events with a wide range variation, for each hour of a day over one year~\cite{garnett2010a}. In addition, physical models for PV forecasting need a large amount of accurate equipment data that are hard to obtain due to measurement and simulation errors~\cite{12345}. On the other hand, AI techniques, such as SVM and ANNs, are solely based on historical statistical data for training~\cite{PINEL20113341}. GPR exploits the advantages of both methods in the sense that it uses both historical data and data fitting approaches to build a robust model \cite{ALHOMOUD2001421}. It should be noted that mathematical modelling of the uncertainties of output as a function of uncertainties of input is outside the scope of this work (see, e.g., \cite{Rasmussen06gaussianprocesses}).

In this work, we propose a probabilistic solar output forecast model using GPR with a Mat\'ern 5/2 kernel function. First, we partition the data into four groups based on time and solar output with k-means clustering. Each hour of the day is considered to be in one specific cluster if it is closer to that cluster's centroid than the three other centroids based on the Euclidean distance. Next, in order to improve the accuracy of the forecast and reduce computational complexity we perform a correlation study to identify the features that have a high impact on solar output. The features selected are: direct solar irradiance, diffused solar irradiance, horizontal solar irradiance, temperature, zenith, and azimuth. We use GPR to relate solar output generation with the selected features and train each cluster using Mat\'ern 5/2 as a kernel function for the forecasting model. To validate the proposed framework we apply it to different datasets from different sites, i.e., Denver, New York, Dallas, San Francisco, and St. Lucia. We validate the results by utilising both k-fold cross validation and holding-out data techniques. We choose 30 random days from the dataset as representatives of different weather conditions as hold-out test data. 

The remainder of the paper is organised as follows. In Section~\ref{Proposed Framework} the data processing is described. More specifically, in Section~\ref{Clustering} the clustering of the dataset is described and in Section~ \ref{correlation} the correlation study to identify the features that have a high impact on solar output is presented. In Section~\ref{sec:prop_fr} the proposed framework is developed. In particular, in Section~\ref{Regression model and kernel selection} the GPR with Mat\'ern 5/2 as a kernel function that relates the solar output with the input features is discussed and in Section~\ref{frame_valid} the framework validation methodologies are presented. In Section~\ref{numerical results}, we illustrate the proposed methodology through five different datasets. In Section~\ref{Conclusion}, we summarise the results and make some concluding remarks.

 \section{Data Processing} 
 \label{Proposed Framework}
 
 In this section, we present the processing that needs to be performed to the data in order to formulate the proposed framework. In particular, we describe the clustering and feature selection methodologies.

 \subsection{Data Clustering} 
 \label{Clustering}

Clustering is an unsupervised pattern classification learning technique used to partition data with high similarity into different groups based on a distance or dissimilarity function \cite{8724682,LIU2019401, 6753773}. The key concepts and different clustering algorithms are discussed in \cite{Jain:1999:DCR:331499.331504}. k-means is a very popular clustering algorithm which is used to cluster data into different groups while each point belongs to a cluster with the least Euclidean distance to the centroid \cite{HUANG20161, Han:2011:DMC:1972541}. In previous studies, k-means is not employed to cluster output solar energy based on time \cite{8477236}, while in our proposal the dataset is clustered based on output and time. PV output has a huge amount of scattering across both the time of day and the day of year. 

 k-means aims to partition the data into $K$ categories in a way that the sum of squares from points to the assigned cluster centres is minimised. In each cluster, all cluster centres are at the mean of the data points which belong to the corresponding cluster. 
 Consider a set $\mathscr{X} = \{x_1,x_2,\dots,x_N\}$ with $N$ elements, where $x_i \in \mathbb{R}^n$ for all of $i = 1,\dots, N$; the data point cluster number $C(i) \in \{1,\dots,K\}$, $i \in \{1,\dots,N\}$; the cluster centroid for cluster $k$ $c_{k} \in \mathbb{R}^n$, $k = 1, \dots, K$; and the Euclidean distance $d(x_i,c_k) = ||x_i- c_k||$, which is the distance between $x_{i}$ and cluster centroid $c_k$.  Then k-means clustering tries to minimize the following squared error function:
\begin{equation}
\label{eq:kmean}
\underset{  \{ c_k\}_{k = 1}^K  }{ \mathrm{minimize}}  \sum_{k=1} ^{K} N_k \sum_{C(i)=k} d^2(x_{i},c_k),
\end{equation}
\noindent where $N_k$ is the number of points assigned to cluster $k$.

\begin{figure}[t!]
\centering
\includegraphics[width=0.45\textwidth]{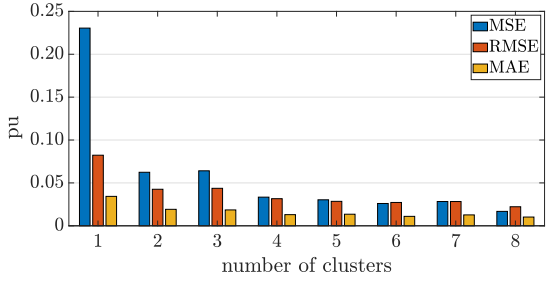} 
\caption{Sensitivity study on the number of clusters by comparing different normalised error metrics values.}
\label{Fig:Data2}
\vspace{-\baselineskip}
\end{figure}

\begin{figure}[b!]
\vspace{-\baselineskip}
\centering
\includegraphics[width=0.45\textwidth]{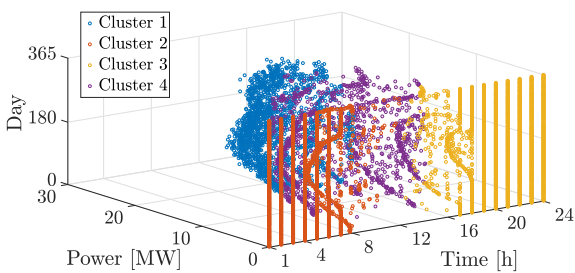} 
\caption{3D graph of four clusters. Different colours represent different clusters.}
\label{Fig:Data1}
\end{figure}

In our framework, the k-means clustering algorithm is used to group the data based on time of day and power output. In order to determine the number of clusters we perform a sensitivity study and compare the increase in the accuracy against the increase in the number of clusters. More specifically, we select solar data from Denver International Airport PV, i.e., Site A (see Table~\ref{site} for more details) and cluster the data into one to eight clusters. For each of the clusters we train a GPR model, as discussed in Section~\ref{sec:prop_fr}, and depict the error between the forecasted and the actual values in~Fig.~\ref{Fig:Data2}. Let us denote by $y_\star^{(t)}$, the forecasted value for solar generation at time $t$, and by $\tilde{y}^{(t)}$ the actual value at time $t$; the error metrics are calculated as follows:
\begin{equation}\label{eq:rmse}
\mathrm{RMSE}=\sqrt{\frac{1}{T_\star}\sum_{t=1}^{T_\star}{\Big(\tilde{y}^{(t)}-y_\star^{(t)}\Big)^2}},
\end{equation}
\begin{equation}\label{eq:mae}
\mathrm{MAE}={\frac{1}{T_\star}\sum_{t=1}^{T_\star}{\left\lvert \tilde{y}^{(t)}-y_\star^{(t)}\right\rvert}},
\end{equation}
\begin{equation}\label{eq:mse}
\mathrm{MSE}={\frac{1}{T_\star}\sum_{t=1}^{T_\star}{\Big(\tilde{y}^{(t)}-y_\star^{(t)}\Big)^2}},
\end{equation}
\noindent where $T_\star$ is the number of hourly intervals we are predicting the solar output. We may also compute normalised values of the above metrics. These metrics, compare how good the prediction of PV is with respect to the number of clusters. Each bar shows the average prediction error for each experiment with different number of clusters. A choice of a large number of clusters increases the computational complexity of the forecasting algorithm since one GPR model needs to be trained for each distinct cluster. From the graph we can see that there is big decrease in all error metrics when the number of clusters is four. However, we notice that after increasing the number of clusters from four to eight there is a marginal decrease in the error metrics.  Thus, the data are grouped into four clusters, which are depicted in Fig. \ref{Fig:Data1}. Clusters 2 and 3 represent early morning and night times. Clusters 1 and 4, represent seasonal variations.  


 \subsection{Features' Selection }
 \label{correlation}
 
After clustering the data into four clusters, we carry out a correlation study to identify the features which are highly related to the output power. There are two different ways to calculate the correlation coefficient: (i) Spearman, which measures only the monotonic correlation between parameters; and (ii) Pearson, which measures the linear relation between power output and each individual feature \cite{8653358}. Studies have found that meteorological data such as temperature and solar irradiance are the main features which affect the solar output\cite{duffie2013solar}. 

We wish to determine the features that affect the solar generation output to the greatest extent. To this end, we calculate the correlation coefficients of solar generation and meteorological features of different datasets from Denver, New York, Dallas, San Francisco, and St. Lucia that may be found in National Solar Radiation (NSR), Iowa Environmental Mesonet (IEM), and  National Renewable Energy Laboratory (NREL) databases. Let us assume, we have a collection of data over a period of $T$ hours. We use the Pearson correlation, which linearly measures the relation between solar output and each feature, defined as a correlation coefficient. The Pearson correlation coefficient between vectors $a \in \mathbb{R}^T$ and $b \in \mathbb{R}^T$ is calculated by the following formulation \cite{10.2307/2331838}:
\begin{equation}\label{eq:correlation}
\rho(a,b)=\frac{1}{T-1}\sum_{t=1}^{T} \frac{{a_{t}- \mu_{a}}}{\sigma_{a}} \cdot \frac{b_{t}- \mu_{b}}{\sigma_{b}},
\end{equation}
\noindent where $a_t$ ($b_t$) is the value of vector $a$ ($b$) at time $t$, $\mu_{a}$ and $\sigma_{a}$ ($\mu_{b}$ and $\sigma_{b}$) are respectively the mean and the standard deviation of $a$ ($b$).

We assume that we have data for different features for $T$ time intervals denoted as $X_{i} \in \mathbb{R}^T$ where $i=1, \dots,M$ is the index of each feature we perform the correlation study on and $Y \in \mathbb{R}^T$ is the time-series solar output. More specifically, meteorological weather data refer to direct solar irradiance, diffused solar irradiance, horizontal solar irradiance, temperature, sky cloud covering, zenith (angle between sun and zenith), azimuth (angle between sun and the North), and albedo, i.e., $M=8$. We calculate the correlation coefficients $\rho(X_i,Y)$ for $i = 1,\dots, M$ to determine which features affect in a greater extent the solar output. 

\begin{table}[t!]
\begin{center}
\begin{tabular}{|c| c|} 
 \hline
Feature & Correlation coefficient  \\ 
 \hline
Direct solar irradiance & 0.71\\
Diffused solar irradiance & 0.64\\
Horizontal solar irradiance &0.81 \\
Temperature & 0.27  \\ 
Zenith  & -0.81 \\ 
Azimuth & 0.49  \\
Sky cloud covering & -0.03\\
Albedo & -0.07 \\
 \hline
\end{tabular}
\end{center}
\caption{Correlation coefficients between solar output and different features.}
\label{tab:corre}
\vspace{-\baselineskip}
\end{table}

 \begin{table}[b]
 \vspace{-\baselineskip}
\centering
\begin{tabular}{|c|c|}
\hline
Cloud coverage & Value in oktas \\
\hline
No clouds & 0 \\ 
Few clouds & {1-2}  \\ 
Scattered clouds & {3-4}  \\ 
Broken clouds & {5-7}  \\ 
Full cloud coverage & 8  \\ 
Sky is hidden from view & 9   \\ 
 \hline
\end{tabular}
\caption{Cloud coverage categories.}
\label{tab_cloud}
\end{table}

In Table \ref{tab:corre} the correlation coefficient values for all attributes are presented. As seen in this table, the value of correlation coefficients for sky cloud covering and albedo are very small in comparison to other features, therefore albedo was eliminated from our feature sets. Although cloud covering has a very small correlation with solar output generation, based on National Aeronautics and Space Administration (NASA) \cite{nasa} research, the amount of sunlight that reaches the Earth can be calculated by using cloud coverage data. In this regard, cloud covering affects the temperature and the amount of sunlight that reaches the Earth. We include cloud covering as a feature to increase the interpretability of the model, i.e., the effect of amount of clouds in the sky is included since it is more understandable for a human observer rather than other measurements, e.g., zenith. 

In weather studies, the sky is categorised into six groups depending on the amount of clouds that are present. Generally, cloud coverage is reported as the number of \textit{oktas}, which is a measurement unit that stands for the amount of clouds in the sky ranging between 0 to 9 \cite{wikicloud}. In Table~\ref{tab_cloud} the different cloud coverage categories and value in oktas are given~\cite{wikicloud}. 

To sum up, based on the aforementioned analysis seven features were identified as the parameters which affect more prominently PV output generation, namely, direct solar irradiance, diffused solar irradiance, horizontal solar irradiance, temperature, zenith, azimuth, and sky cloud covering. The number of features whose relationship with solar generation was originally studied was eight. Even though this was a small change; the accuracy of the forecasts was better and the computational complexity was reduced in the case of seven selected features compared to eight.  
 
\section{Proposed Framework}
\label{sec:prop_fr}

In this section, the stochastic framework for the short-term forecast  of PV output is presented. More specifically, the formulation of the GPR is described and the validation methodologies are discussed. 

 \subsection{Gaussian Process Regression}
  \label{Regression model and kernel selection}
 
 In this work, a model is trained for each cluster using a GPR model which is a supervised learning technique. In supervised learning, we aim to learn a mapping function that relates the input feature set data to the output data. In fact, GPR is a kernel based nonlinear nonparametric regression technique, in which the covariance function plays a crucial role in defining the relation between input data and the responses. 
 
 Let the training set $\mathscr{S}= \{(x^{(t)},y^{(t)})\}_{t=1}^T$ be a set of i.i.d. samples from some unknown distribution, where $T$ is the period of available data with one hour resolution; $q$ stands for the number of selected features, i.e., $q=7$; $x^{(t)} \in\mathbb{R}^q$ is the vector containing all selected features at time $t$; and $y^{(t)} \in \mathbb{R}$ the solar output at observation $t$. With the use of a Gaussian model we may relate the input with the output terms by:
\begin{equation} 
\label{eq:GPR}
y^{(t)}=f(x^{(t)}) + h{(x^{(t)})}^{\top}\beta+\epsilon^{(t)}, \text{ for } t = 1,\dots, T,
\end{equation}
where $\epsilon^{(t)}$ are i.i.d. ``noise'' variables with independent $\mathscr{N}(0, \sigma^2)$ distributions, $f(x^{(t)})$ is the mapping function $\mathbb{R}^q \rightarrow \mathbb{R}$ and $h{(x^{(t)})}$ is a set of a fixed basis function. The explicit use of basis functions is a way to specify a non-zero mean over $f(x^{(t)})$. In this work we assume that $h{(x^{(t)})}$ is a $q \times 1$ vector whose all entries are equal to the constant value of one, and $\beta$ is the basis function coefficient $q\times 1$ vector and is evaluated by maximising a likelihood function as described below. For notational convenience, we define:
\begin{equation*}
X = \begin{bmatrix} (x^{(1)}) \\ \vdots  \\(x^{(T)}) \end{bmatrix} \in \mathbb{R}^{T \times q}, y = \begin{bmatrix} y^{(1)} \\ \vdots  \\y^{(T)} \end{bmatrix} \in \mathbb{R}^{T}, \epsilon = \begin{bmatrix} \epsilon^{(1)} \\ \vdots  \\\epsilon^{(T)} \end{bmatrix} \in \mathbb{R}^{T}, 
\end{equation*}
\begin{equation*}
f = \begin{bmatrix} f(x^{(1)}) \\ \vdots  \\f(x^{(T)}) \end{bmatrix} \in \mathbb{R}^{T}, H = \begin{bmatrix} h(x^{(1)}), \dots, h(x^{(T)}) \end{bmatrix}=\mathbb{1}_{q \times T} , 
\end{equation*}
\noindent where $\mathbb{1}_{q \times T}$ is a $q$ by $T$ matrix whose all elements are one. In matrix form we may rewrite \eqref{eq:GPR} as
\begin{equation}
\label{eq:gpr_matrix}
y=f(X) + H^{\top}\beta+\epsilon.
\end{equation}
\noindent We assume a prior distribution over functions $f(X)$ as 
\begin{equation}
f(X) \sim \mathscr{N}(0, K(X, X)),
\end{equation}
\noindent where $0$ is the mean value; $K(X,X)$ is the covariance matrix:
\begin{equation*}
K(X,X) = \begin{bmatrix} 
k(x^{(1)}, x^{(1)}) & \dots & k(x^{(1)}, x^{(T)}) \\
 \vdots & \ddots & \vdots \\
  k(x^{(T)}, x^{(1)}) & \dots &k(x^{(T)}, x^{(T)})
  \end{bmatrix},
\end{equation*}
\noindent where $k(\cdot,\cdot)$ is the covariance or kernel function. By using the kernel function we aim to actively model the unknown relationship between the input and the output variables. The kernel function is defined based on the likely pattern that we can observe in the data. One assumption to model the kernel may be that the correlation between any two points in our input set, i.e., $x^{(t)},x^{(t')} \in \mathscr{S}$, with $t, t' = 1,\dots, T, t \neq t'$, decreases with increasing the euclidean distance between them. This means that points with similar features behave similarly. Under this assumption, in this work we use the Mat\'ern 5/2 as a kernel function, which is parameterised as follows:
\begin{eqnarray}\label{eq:linearreg11113}
k(x^{(t)},x^{(t')})&= \sigma^2_{f} \left(1+\frac{\sqrt{5}d(x^{(t)},x^{(t')})}{\sigma_l}+\frac{5d^2(x^{(t)},x^{(t')})}{3\sigma^2_l}\right) \nonumber \\
&e^{-\frac{\sqrt{5}d(x^{(t)},x^{(t')})}{\sigma_l}},
\end{eqnarray}
\noindent where $d(x^{(t)},x^{(t')})$ is the euclidean distance between any two input observations $x^{(t)},x^{(t')} \in \mathscr{S}$ as defined in Section~\ref{Clustering}; 
$\sigma_l$ and $\sigma_f$, are two other kernel parameters which show respectively the characteristic length scale and the signal standard deviation that both belong in $\mathbb{R}^{q}$. The characteristic length scale $\sigma_l$ defines how far the response variable $y^{(t)}$ needs to be away from the predictor $x^{(t)}$ to become uncorrelated. These two parameters are greater than zero and are formulated as follows:
\begin{equation}
\sigma_l = 10^{\theta_l}, \sigma_f = 10^{\theta_f}.
\end{equation}
\noindent We now define a new parameter $\theta$ to be:
 \begin{equation}
\label{eq:linearreg113}
\theta= \begin{bmatrix} \theta_l\\ 
\theta_f \end{bmatrix} = \begin{bmatrix} \log(\sigma_l)\\ 
\log(\sigma_f) \end{bmatrix}\in \mathbb{R}^{q\times2}  .
\end{equation}

From \eqref{eq:gpr_matrix} we may write that 
 \begin{equation}
 \label{eq:linearreg2}
y|f(X),X \sim \mathscr{N}(H^{\top}\beta,\sigma^2I+K(X,X)),
\end{equation}
since both $f(X)$ and $\epsilon$ have zero means. In order to determine the distribution that $y$ follows, we need to determine three parameters, i.e., $\beta$, $\sigma^2$ and $\theta$. $K(X,X)$ is a function of $\theta$ as may be seen in \eqref{eq:linearreg11113}-\eqref{eq:linearreg113}. $\beta$, $\sigma^2$, and $\theta$ are also known as the hyperparameters of the kernel function. In order to estimate the parameters we maximise the following marginal log-likelihood function
 \begin{equation}
 \label{eq:kxx}
\mathrm{log} P(y|f(X),X)= \mathrm{log} P(y|X, \beta, \theta, \sigma^2).
\end{equation}
 \noindent Thus, the estimates of $\beta$, $\theta$, and $\sigma^2$ denoted by $\hat{\beta}$, $\hat{\theta}$ and $\hat{\sigma}^2$ are given by 
 \begin{equation} 
 \label{hyperpara}
     \hat{\beta}, \hat{\theta}, \hat{\sigma}^2=\underset{\beta,\theta, \sigma^2}{ \mathrm{arg max}} \, \mathrm{log} P(y| X,\beta, \theta, \sigma^2).
 \end{equation}
 We may write from \eqref{eq:linearreg2} and \eqref{eq:kxx} that 
  \begin{equation}\label{eq:kxxxx}
P(y|X)=P(y|X, \beta, \theta, \sigma^2)=\mathscr{N}(H^T\beta, K(X,X)+\sigma^2I).
\end{equation}
\noindent Thus, the marginal log-likelihood  function is
\begin{equation} 
\label{logliklihood}
\begin{split}
\mathrm{log} P(y| X,\beta, \theta, \sigma^2)=  -\frac{1}{2} (y-H^\top \beta)^T [K(X,X)+\sigma^2I]\;^{-1}\\
(y-H^\top \beta)- \frac{1}{2} \textrm{log} \; 2\pi -\frac{1}{2} \mathrm{log} |K(X,X)+\sigma^2I |.
\end{split}
\end{equation}
We concentrate the likelihood function for the subset of parameters, $\sigma^2$ and $\theta$, by expressing $\beta$ as a function of the parameters of interest and replacing them in the likelihood function. Thus, we have that the estimate of $\beta$ for given $\theta$  and $\sigma^2$ is:
\begin{equation} \label{eq:beta}
  \begin{split}
      \hat{\beta}(\theta,\sigma^2)= [H^\top[K(X,X|\theta)+\sigma^2 I]\;^{-1}H]\;^{-1}\\
      H^\top[K(X,X|\theta)+\sigma^2I]\;^{-1}y.
  \end{split}  
\end{equation}
By substituting \eqref{eq:beta} in \eqref{logliklihood} we have 
\begin{equation} \label{logliklihood1}
\begin{split}
\mathrm{log} P(y| X,\hat{\beta}(\theta,\sigma^2), \theta, \sigma^2)=  -\frac{1}{2} (y-H\hat{\beta}(\theta,\sigma^2))^T \\
 [K(X,X|\theta)+\sigma^2I]\;^{-1}
(y-H\hat{\beta}(\theta,\sigma^2)) \\- \frac{1}{2} \mathrm{log} \; 2\pi -\frac{1}{2} \mathrm{log} |K(X,X|\theta)+\sigma^2I|.
\end{split}
\end{equation}
We now may determine the hyperparameters as the output of the above optimisation problem.

Once the hyperparameters are evaluated we may use \eqref{eq:linearreg2} to predict the output of solar generation based on the input parameters. More specifically, $\{x_*^{(t)}\}_{t=1}^{T_*}$ be a set of i.i.d. input points of the features drawn from the same unknown distribution; we will plug these values in \eqref{eq:linearreg2} and the unknown $\{y_*^{(t)}\}_{t=1}^{T_*}$ can be calculated as the predicted solar output value for the time period $T_*$. More details on GPR model may be found in~\cite{Rasmussen06gaussianprocesses, Schulz095190}.

After training our data and estimating the kernel parameters for each of the four clusters we can use the proposed framework for solar generation forecasting. 

 \begin{figure}[t]
     \centering
  \includegraphics[width=.42\textwidth]{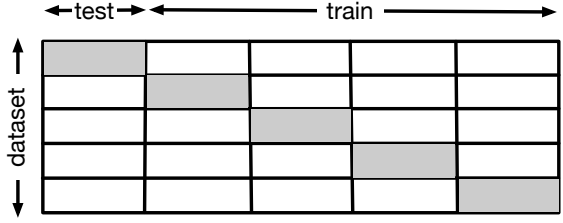}   
  \caption{Diagram of k-fold cross-validation with k=5; the grey boxes refer to testing the data and the white to training.}
     \label{fig_val}
     \vspace{-\baselineskip}
 \end{figure}

\subsection{Framework Validation}
 \label{frame_valid}
 
To test the accuracy of the proposed method for solar output forecasting, different tests and validation methods are exploited. k-fold cross-validation and hold out validation are the most prevalent test methods used in recent studies \cite{James:2014:ISL:2517747}. As depicted in Fig.~\ref{fig_val}, in k-fold cross validation the whole data set is split into k folds: at each time, k-1 folds are used as training set and a one-fold as testing set, until all folds used to build the forecast model, typical values for k range between 3 to 10~\cite{wiki:xxx}.  In addition, hold-out is used to avoid overfitting \cite{7544814}. In this work both methods are used for test and validation. 

In our implementation, 30 days of a year are randomly selected as hold-out data while the remaining data are used for training and testing using 5-fold cross validation.

\section{Numerical results} \label{numerical results}

In this section, results of the proposed framework at five sites are given as well as comparisons with existing forecasting methodologies in the literature. In Section~\ref{dataset}, the five sites' information is given; in Section~\ref{One dataset results and analysis} detailed results and analysis of site A are given so that the reader better understands the proposed framework. In Section~\ref{all datasets results and analysis}, summarised results for all sites are given as well as comparisons with other methods to prove the efficiency of the proposed framework. 

\subsection{Dataset Information}
 \label{dataset}
To test the efficiency of the model, different datasets from different sites are used based on available historical data from National Solar Radiation, Iowa Environmental Mesonet (IEM) and National Renewable Energy Laboratory. The five sites' details are given in Table~\ref{site}.

 \begin{table}[t]
\centering
\resizebox{0.49\textwidth}{!}{
\begin{tabular}{|c|c|c|c|c|}
\hline
Site & Location & Size [MW] & Latitude [$^\circ$] &  Longitude [$^\circ$] \\
\hline
A & Denver Intl Airport &30 &39.8561 N &104.6737 W \\
B & John F. Kennedy IntlAirport & 30 & 40.6413 N & 73.7781W \\
C & Dallas Executive Airport  & 35 & 32.6807 N &96.8672 W \\
D  &San Francisco Intl Airport & 30 &37.6213 N & 122.3790 W\\
E & St Lucia & 0.433 & 27.498 S & 153.013 E \\
 \hline
\end{tabular}}
\caption{Site description.}
\label{site}
\end{table}

\begin{figure}[b]
    \vspace{-\baselineskip}
     \centering
  \includegraphics[width=.50\textwidth]{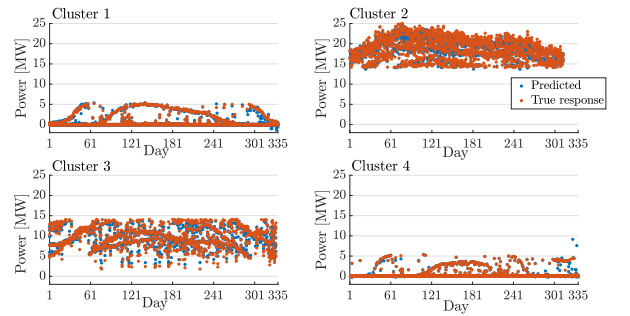}   
  \caption{Proposed framework predictions of the training data set.}
     \label{fig:trainingresults}
 \end{figure}

\subsection{Framework Implementation on Site A} 
\label{One dataset results and analysis}

The model is tested on the Denver International Airport PV plant, i.e., site A. The dataset comprises of hourly attributes' values from 2006, i.e., of 8760 data points for each feature and the solar generation output. The experiments are run 30 independent times for each cluster; 30 random days are selected as hold-out data that are representative of different days of the year during different seasons. As described in Sections \ref{Proposed Framework} and \ref{sec:prop_fr}, the training set is partitioned into four clusters and all clusters are trained using GPR Mat\'ern 5/2 and tested using 5-fold cross validation and hold-out methods. 

\begin{figure}[t]
  \vspace{-\baselineskip}
     \centering
  \includegraphics[width=.50\textwidth]{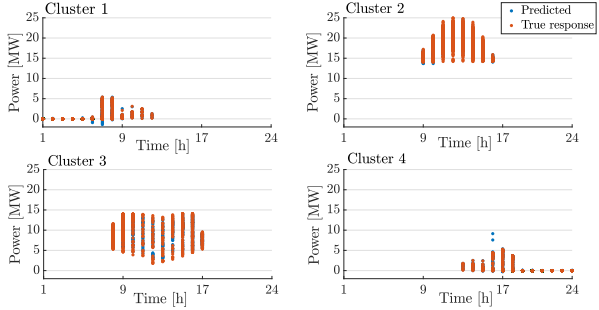}   
  \caption{Proposed framework hourly predictions of the training data set.}
     \label{fig:hourtrainingresults}
     
 \end{figure}

We first train the GPR model with the available hourly dataset of 335 (365-30=335) days and depict the predictions of the training data set for the four different clusters in Fig.~\ref{fig:trainingresults}. We use 5-fold cross validation as a test and validation method for our training set which comprise hourly data-points of 335 days. As seen in Fig.~\ref{fig:trainingresults}, clusters represent the hourly points for 335 day of training dataset. Since clusters are partitioned based on similarities between the points, as shown, each cluster follows specific patterns which prove the similarity of the data in them. Moreover, for each cluster a remarkable proximity between the actual data and predicted values is seen.  
In order to further understand the value of clustering in Fig. ~\ref{fig:hourtrainingresults} the hourly predictions of the testing data are depicted. It may be seen that there are hours which belong to, e.g., three clusters, that means that even the same hour patterns may be different based on which cluster they are identified to be in. In other words, different hours in different days even if the days are in the same season, may behave completely different. It may also be seen in Fig.~\ref{fig:hourtrainingresults} that Clusters 2 and 3 represent early morning and night times and Clusters 1 and 4, represent seasonal variations, as also mentioned in Section~\ref{Clustering}.

 \begin{table}[b]
\centering
\begin{tabular}{|c |c| c |c |c|} 
 \hline
Site A & RMSE (MW) & MAE (MW) & RMSE[\%] & MAE[\%]  \\ 
 \hline
Training set & 1.24 & 0.36 & 4.18 & 1.22 \\ 
Test set  & 1.23 & 0.56 & 4.12 & 1.89\\ 
 \hline
\end{tabular}
 \caption{Site A forecasts' error metrics.}
  \label{tab:DEN}
   \vspace{-\baselineskip}
\end{table}

\begin{figure}[t]
     \centering
  \includegraphics[width=.5\textwidth]{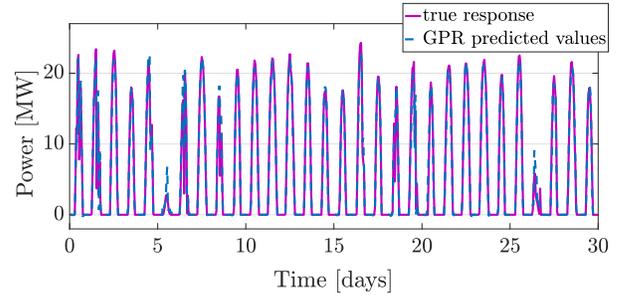}   
  \caption{1-24 hour ahead prediction for 30 random days.}
     \label{fig:onemonthpredict}
     \vspace{-\baselineskip}
 \end{figure}

\begin{figure}[b]
  \vspace{-\baselineskip}
     \centering
  \includegraphics[width=.5\textwidth]{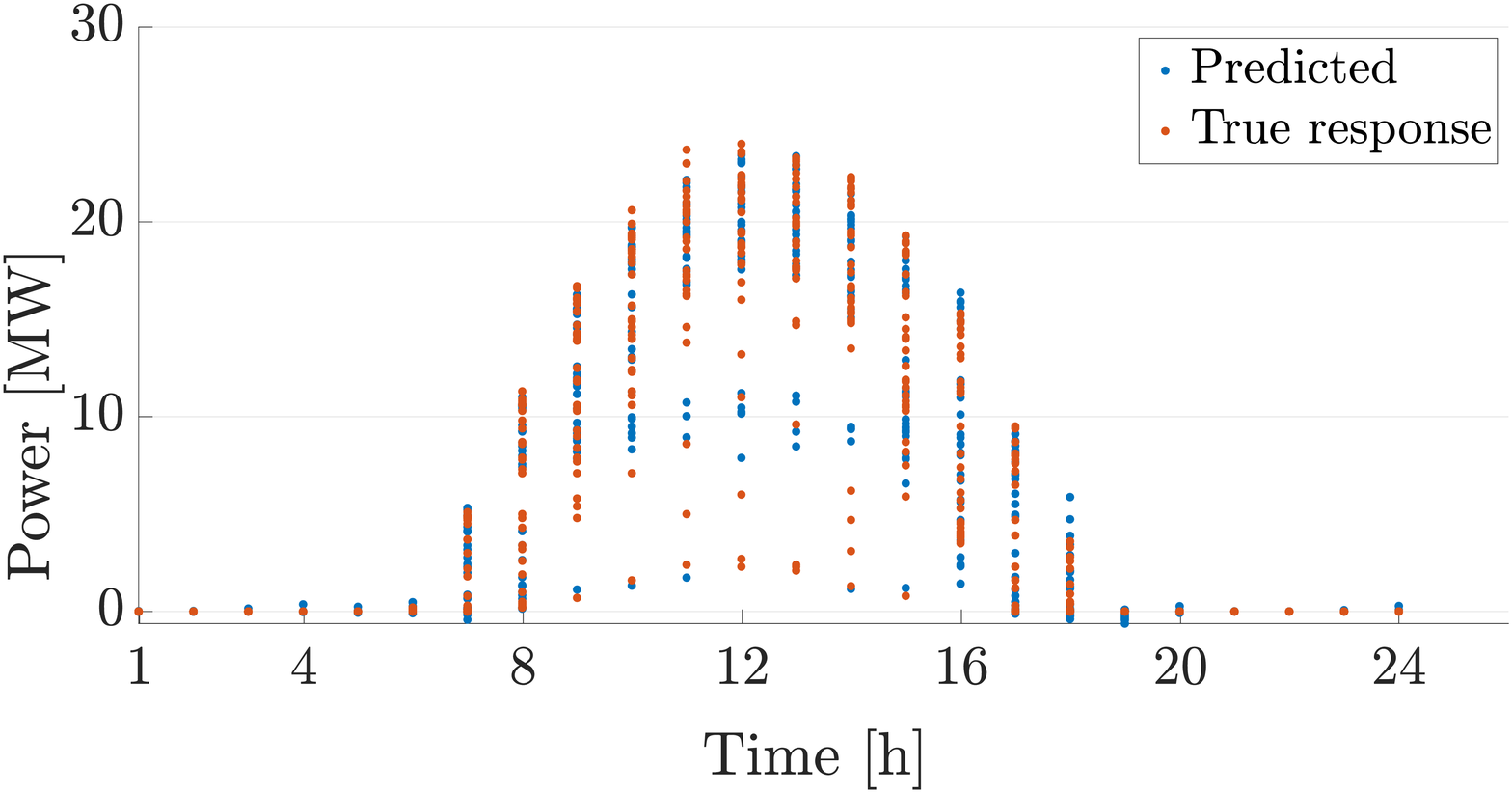}   
  \caption{30 random days 24-hour prediction with one hour intervals, Denver.}
     \label{fig:30daysprediction}
 \end{figure}

In Fig.~\ref{fig:onemonthpredict} the forecasts and actual values of the 30 hold-out selected days, that are representative of different seasons, are depicted. The x-axis of the figure has 30 days, that correspond to 24-hour intervals for each day. As it may be seen the two values are very close to each other. Another visual representation of the same result, i.e., the daily hourly forecast of the 30 days, may be seen in Fig.~\ref{fig:30daysprediction}, where we notice that the predicted and the actual values follow the same pattern.

In order to test the accuracy of the forecasts we use the following statistical metrics: RMSE, and MAE, as defined in \eqref{eq:rmse} and \eqref{eq:mae}. The statistical results for the training set and the test set are summarised in Table \ref{tab:DEN}. The error metrics of the testing data between the actual data and the predicted values are based on the average error of all 5 folds for the training set. To interpret these values, notice that the higher RMSE and MAE values, the less predictive the model is. It should be noted that the test results are expected to be different from the training set results, since 30 hold-out days are not shown to the model during the training process. However, the results with any test set should be approximately the same as those obtained with the training set, as it may be seen in Table \ref{tab:DEN}.

  \begin{figure}[t]
     \centering
  \includegraphics[width=.5\textwidth]{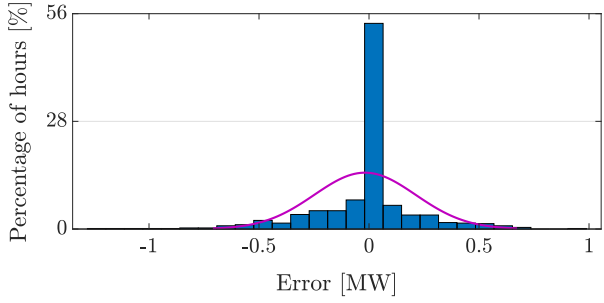}   
  \caption{Probability distribution fitting of average error of one sample model.}
     \label{fig:errordist}
 \end{figure}

The error between the actual and the forecasted value for the 30 hold-out days is depicted in Fig.~\ref{fig:errordist}. The average prediction error of the hold-out days for one cluster is fitted into a normal distribution. In this figure, y-axis represents the percentage of hours; it may be seen that 54\% of the hours, i.e, 382 hours, that the prediction error was less than 0.03 MW. In order to provide a confidence certificate to the forecast we use the confidence interval (CI)~\cite{8096334}. The selection of a confidence level for an interval determines the probability that the confidence interval produced will contain the true parameter value. Common choices for the CI are 0.90, 0.95, and 0.99. The CI is defined as follows:
\begin{equation}
\label{eq:CI}
\mathrm{CI}=\left(\Bar{\epsilon}-z^* \frac{\sigma_\epsilon}{\sqrt{T_\star}},\Bar{\epsilon}+z^* \frac{\sigma_\epsilon}{\sqrt{T_\star}} \right),
\end{equation}
\noindent where $\bar{\epsilon}$ is mean value of the errors, $\sigma_\epsilon$ is the standard deviation and $T_\star$ is the sample size of the errors. The value $z^\star$ represents the point on the standard normal density curve such that the probability of observing a value greater than $z^\star$ is equal to $p$. The relationship between $\mathrm{CI}$ and $p$ is $p=(1-\mathrm{CI})/2$. Thus, if we wish to have a $\mathrm{CI}$ of 95\% then $p = 0.025$. The value $z^\star$ such that $P(Z >z^\star) = 0.025$, or $P(Z < z^\star) = 0.975$, is equal to $1.96$ as we may find in a standard normal distribution table. As the level of confidence decreases, the size of the corresponding interval will decrease. By fitting a normal distribution in Fig.~\ref{fig:errordist} we have mean value $\bar{\epsilon} = 0.03$  and a standard deviation $\sigma_\epsilon =0.50 $. Now, we may calculate the CI for various confidence levels; for instance with $95\%$ confidence level, the difference between the actual data and the prediction value of each point ranges between $[-0.47,0.43]$ MW or $[-1.6\%,1.4\%]$.

\subsection{Summary of results of all sites} 
\label{all datasets results and analysis}

In order to further validate our framework we have applied it to the remaining four sites as given in Table~\ref{site}. Following the same procedure as described in more details in Section~\ref{One dataset results and analysis} we have four clusters per site; from each dataset we hold-out 30 representative days and train with the remaining data a GPR model for each cluster. The results for each site are summarised in Table~\ref{tab:trainresults} for the training dataset and in Table~\ref{tab:testresults} for the test dataset.

 \begin{table}[b]
\centering
\begin{tabular}{|c| c| c| c| c|} 
 \hline
Site & RMSE [MW] & MAE [MW] & RMSE [\%] & MAE [\%] \\ 
 \hline
A & 1.25 & 0.36 & 4.18 & 1.22 \\ 
B  & 1.39 & 0.63 & 4.18 & 1.91\\ 
C & 1.51 & 0.59& 4.33 & 1.69\\  
D  &  1.39 & 0.30 & 4.21 & 0.92\\  
E &0.019 &0.008 &4.48 &1.96 \\
 \hline
 \end{tabular}
  \caption{Training set error metrics for all sites. }
 \label{tab:trainresults}
 \end{table}

 \begin{table}[b]
\centering
\begin{tabular}{|c| c| c| c| c|} 
 \hline
Site & RMSE [MW] & MAE [MW] & RMSE [\%] & MAE [\%] \\ 
 \hline
A &  1.23 & 0.56 & 4.12 & 1.89\\ 
B  & 1.51 & 0.66 & 4.58 & 2.00\\ 
C  & 1.61 & 0.72& 4.60 & 2.06\\ 
D  & 1.44 & 0.65 & 4.38 & 1.98\\ 
E & 0.015 &0.008& 3.48 &1.85\\
 \hline
\end{tabular}
\caption{Test set error metrics for all sites.}
 \label{tab:testresults}
\end{table}
As seen above, the results for all datasets are approximately in the same range, which means that the model may be applied in any site under the assumption that the data of the selected features are available. 

To further prove the effectiveness of the proposed framework we compare our results with other recent studies. In order to make the comparison meaningful, we need to have access to the same set of data. The work presented in \cite{7285739, Golestaneh} use the same data for site E, which are available from University of Queensland. The temporal resolution of the data in~\cite{7285739} is 1-minute; however since we are interested in hourly values we select historical data with hourly resolution. We used 2012 data for training and 2013 data for testing. The authors in~\cite{7285739} calculate predictions for each of the four seasons, i.e., fall, winter, spring and summer. In~\cite{Golestaneh} the authors calculate hourly forecasts using their proposed ELM method and the traditional feed-forward back propagation neural network (FFBPG). Yearly results are better than each season prediction in \cite{7285739, Golestaneh}, as may be seen in Table~\ref{tab:111}. 

The errors of the proposed framework are small since the variation of solar output over different times of day and year is taken into account with the use of k-means clustering. The use of a clustering algorithm results in similar points that belong in the same cluster being trained with a GPR model. More specifically, k-means divides similar data in one group which follows a distribution with specific characteristics which makes training of each cluster more efficient with lower errors. Moreover, the use of an appropriate kernel function that relates the input features to output, improve the forecast. In this work using Mat\'ern 5/2 as a kernel function increase the accuracy of forecast due to the capability of the kernel in solving stochastic problems.


 \begin{table}[t]
\centering
\begin{tabular}{|c| c | c | c|} 
 \hline
\multicolumn{2}{|c|}{}& RMSE[\%] & MAE[\%]  \\ 
 \hline
 \multicolumn{2}{|c|}{Proposed framework} & \textbf{3.48} & \textbf{1.85} \\
 \hline
\multirow{4}{*}{[11]}& Fall  & 13.85 & 8.48 \\
& Winter & 7.67 & 4.16\\
 & Spring &13.6 & 8.08\\
& Summer& 16.43 & 10.73\\
\hline
 \multirow{2}{*}{[42]}&ELM & 12.84 & 6.68  \\
& FFBPG &13.33 & 7.53  \\
 \hline

\end{tabular}
 \caption{Forecast error metrics based on different methodologies for site E.}
 \label{tab:111}
\end{table}

\section{Conclusion} 
\label{Conclusion}

In this paper, we proposed a probabilistic framework for short-term photovoltaic forecasting. Since solar output relies on solar irradiance, we clustered our data in four groups based on day-time. Two clusters represent early morning and night times; and the remaining two represent seasonal variations. After clustering data into four clusters, we carried out a correlation study to identify the features which are highly related to solar output power. The seven selected features that affected more prominently the PV output generation were: direct solar irradiance, diffused solar irradiance, horizontal solar irradiance, temperature, zenith, azimuth, and sky cloud covering. We then trained a model for each of the four clusters using GPR in order to learn the relationship between the seven input features and the PV generation. GPR is a kernel based nonlinear nonparametric regression technique, in which the covariance function plays a crucial role. In this work, we selected the Mat\'ern $5/2$ as a covariance or kernel function. This function was selected under the assumption that the correlation between any two points in the input  feature set decreases with increasing the euclidean distance between them. To test the accuracy of the proposed method for solar output forecasting, different tests and validation methods were exploited, i.e., k-fold cross-validation and hold-out validation methods.

In the case studies, we demonstrated the framework implementation in five different sites. For each site, the experiments were run 30 independent times for each cluster, i.e., 30 random days were selected as hold-out data that were representative of different days of the year during different seasons. The largest RMSE and MAE were 4.60 \% and 2.06 \% respectively, showing the efficacy of the proposed framework. Furthermore, the proposed framework was compared with existing forecasting methodologies and it was found that its predictions were more accurate based on statistical metrics. 

\balance

\bibliographystyle{IEEEtran}
\bibliography{journals-full,references}

\end{document}